\documentclass[envcountsame]{llncs}
\usepackage{amsmath, amssymb}
\usepackage{algorithmic,enumitem}
\usepackage{graphicx,subfig}
\usepackage{url}

\spnewtheorem{algorithm}[theorem]{Algorithm}{\bfseries}{}

\newenvironment{turing}[2]{
	\textsc{#1}$($#2$)$
	\begin{algorithmic}[1]
}{
	\end{algorithmic}
}

\newcommand{\abs}[1]{\left|#1\right|}
\renewcommand{\~}[1]{\widetilde{#1}}
\newcommand{\symdiff}{\vartriangle}
\newcommand{\st}{\,:\,}

\newcommand{\BREAK}{\textbf{break}}
\renewcommand{\deg}{\textit{deg}}

\let\oldsubsection\subsection
\renewcommand{\subsection}[1]{\vspace{-1em}\oldsubsection{#1}\vspace{-0.5em}}

\newenvironment{keywords}{
       \list{}{\advance\topsep by0.35cm\relax\small
       \leftmargin=1cm
       \labelwidth=0.35cm
       \listparindent=0.35cm
       \itemindent\listparindent
       \rightmargin\leftmargin}\item[\hskip\labelsep
                                     \bfseries Keywords:]}
     {\endlist}

\title{A Nearly-Quadratic Gap Between Adaptive and Non-Adaptive Property Testers}
\author{Jeremy Hurwitz\thanks{Supported by NSF grants CCF-0830787 and CCF-0829909}}
\institute{Department of Computing and Mathematical Sciences, California Institute of Technology, Pasadena California 91125 \\
\email{jhurwitz@cs.caltech.edu}}

\begin{document}

\maketitle

\vspace{-1.5em}\begin{abstract}
We show that for all integers $t\geq 8$ and arbitrarily small $\epsilon>0$, there exists a graph property $\Pi$ (which depends on $\epsilon$) such that $\epsilon$-testing $\Pi$ has non-adaptive query complexity $Q=\~{\Theta}(q^{2-2/t})$, where $q=\~{\Theta}(\epsilon^{-1})$ is the adaptive query complexity. This resolves the question of how beneficial adaptivity is, in the context of proximity-dependent properties (\cite{benefits-of-adaptivity}). This also gives evidence that the canonical transformation of Goldreich and Trevisan (\cite{canonical-testers}) is essentially optimal when converting an adaptive property tester to a non-adaptive property tester.

To do so, we provide optimal adaptive and non-adaptive testers for the combined property of having maximum degree $O(\epsilon N)$ and being a \emph{blow-up collection} of an arbitrary base graph $H$.
\end{abstract}
\vspace{-3em}

\section{Introduction}

In this paper, we consider the power of adaptive versus non-adaptive queries for property testers in the dense-graph model. In this model, the algorithm is given access to the input graph $G=([N],E)$ via an oracle $g:[N]\times[N]\rightarrow\{0,1\}$ such that $g(u,v)=1$ if and only if $(u,v)\in E$. A graph is said to be \emph{$\epsilon$-far} from a particular property if at least $\epsilon N^2$ edges must be added or deleted to yield a graph with the desired property.

Given a graph property $\Pi$, we say that a (randomized) algorithm $\mathcal{A}$ is an \emph{$\epsilon$-tester} for $\Pi$ if $\Pr[\mathcal{A}(G)=\textsc{Accept}]>2/3$ for all $G\in\Pi$ and $\Pr[\mathcal{A}(G)=\textsc{Reject}]>2/3$ for all $G$ which are $\epsilon$-far from $\Pi$.

\subsection{Adaptive, Non-Adaptive, and Canonical Testers}

Property testers can be broadly classified into two types according to how the queries are determined. In an \emph{adaptive algorithm}, the results of previous queries can be used when determining the next query. A \emph{non-adaptive algorithm}, on the other hand, determines all of its queries in advance. 

Most of the initial work in the dense-graph model focused on non-adaptive algorithms. In fact, these early algorithms used an even more restricted framework, termed \emph{canonical algorithms} by \cite{canonical-testers}. In a canonical algorithm, the tester chooses a random subset of the vertices and queries the entire induced subgraph.

Given a property $\Pi$ and an error parameter $\epsilon>0$, let $q$ be the adaptive query-complexity, $Q$ be the non-adaptive query complexity, and $\widetilde{Q}$ be the canonical query complexity. By definition, $q\leq Q\leq\widetilde{Q}$. A natural question, then, is to determine the exact relationship between these parameters.

In \cite{canonical-testers} and \cite{combinatorial-characterization}, the authors note that any algorithm which makes $q$ queries considers at most $2q$ vertices. It therefore suffices to query the subgraph induced by a random set of $2q$ vertices and then locally simulate the adaptive algorithm. This \emph{canonical transformation} shows that, for any property, $\widetilde{Q}<2q^2$.

It is easy to see that there exist properties with $\widetilde{Q}=\Omega(Q^2)$,\footnote{Let $\Pi$ consist of only the empty graph. Then $Q=O(\epsilon^{-1})$, while $\widetilde{Q}=\Omega(\epsilon^{-2})$.} thereby showing that the canonical transformation is optimal when converting (adaptive or non-adaptive) algorithms into canonical algorithms. Indeed, \cite{testing-graph-blowup} and \cite{algorithmic-aspects} show that a quadratic performance gap exists between non-adaptive and canonical testers for many natural properties.

Similarly, we can consider the relative power of adaptive versus non-adaptive (but not necessarily canonical) property testers. It is widely believed that the canonical transformation remains optimal, in the worst case, even for this more modest goal. In other words, it is believed that there exist properties such that $Q=\Omega(q^2)$. However, proving such a separation between the adaptive and non-adaptive complexities has proven elusive  -- no unconditional gap was known until Goldreich and Ron demonstrated a property where $Q=\~{\Omega}(q^{3/2})$ in \cite{algorithmic-aspects}.

In light of this difficulty, researchers have considered two modifications of the problem. The first approach, used by \cite{benefits-of-adaptivity}, considers \emph{proximity-dependent} properties which depend on the tolerance parameter $\epsilon$. In particular, they considered the combined property of being bipartite and having degree at most $O(\epsilon N)$, and achieved a gap of $Q=\~{\Theta}(q^{-4/3})$.

The second approach, used by \cite{algorithmic-aspects}, is to consider \emph{promise problems}. In this context, the authors showed that there exist properties such that $Q=\Omega(q^{-2+\delta})$ for all $\delta>0$, exhibiting the first nearly-quadratic gap between adaptive and non-adaptive queries in the dense-graph model. More specifically, the authors demonstrated a hierarchy of gaps of the form $Q=\Theta(q^{-2+2/t})$ for each integer $t\geq 2$. Unfortunately, the promise they use, while natural, is quite strong, and it is currently unclear how to remove the promise.

In this paper, we achieve a nearly-quadratic gap without using a promise, at the expense of making the properties proximity-dependent. This proves that for all $\delta>0$ and arbitrarily small $\epsilon$, there exists a property $\Pi$ (which depends on $\epsilon$), such that $\epsilon$-testing $\Pi$ requires $Q>q^{2-\delta}$ non-adaptive queries, where $q$ is the adaptive query complexity. As in \cite{algorithmic-aspects}, we also strengthen the result and establish a hierarchy of relationships between the adaptive and non-adaptive query complexities.

\begin{theorem}[Main Theorem]\label{theorem:hierarchy-theorem}
For all $t\geq 8$ and arbitrarily small $\epsilon$, there exists a graph property $\Pi$ (which depends on $\epsilon$) such that $\epsilon$-testing $\Pi$ has non-adaptive query complexity $Q=\~{\Theta}(q^{2-2/t})$, where $q=\~{\Theta}(\epsilon^{-1})$ is the adaptive query complexity.
\end{theorem}

Theorem \ref{theorem:hierarchy-theorem} and \cite{algorithmic-aspects} both provide strong evidence that the canonical transformation is optimal in the general case. Although each result technically leaves open the possibility of a better transformation between adaptive and non-adaptive testers, each does so in a different, and very restricted, way. As a result, any such transformation would have to be very unnatural and would have to depend sensitively on the internal structure of the adaptive tester.

\subsection{Graph Blow-Ups and Blow-Up Collections}

A \emph{graph blow-up} consists of replacing each vertex of a graph with a cluster of vertices (a rigorous definition is given in section \ref{section:notation}). This operation is frequently used in studying the dense-graph model (see, for example, \cite{testing-subgraphs-in-large-graphs}, \cite{easily-testable-induced-subgraphs}, and \cite{hierarchy-theorems}).

The complexity of testing whether a graph is a blow-up of a fixed graph $H$ was essentially resolved in \cite{lowest-level-of-query-complexity} (see also \cite{testing-graph-blowup}), where it is shown that, for any $H$, the adaptive query complexity is $O(\epsilon^{-1})$ and the non-adaptive query complexity is $\~O(\epsilon^{-1})$.

A graph is a \emph{blow-up collection} if it can be partitioned into disjoint subgraphs, each of which is a blow-up of $H$. This notion was implicitly introduced in \cite{algorithmic-aspects}, which showed the following lower bound.
\begin{lemma}[\cite{algorithmic-aspects}, Lemma 5.6]\label{lemma:nonadaptive-lower-bound}
Let $H$ be a simple $t$-cycle, with $t\geq 4$. Testing whether $G$ is a blow-up collection of $H$ requires $\Omega(\epsilon^{-(2-2/t)})$ non-adaptive queries, even given the promise that $G$ has maximum degree $2t\epsilon N$.
\end{lemma}

In this paper, we prove tight upper-bounds for both the adaptive and non-adaptive cases. Specifically, we show that a tester can determine whether $G$ is a blow-up collection of any given $H$, given that same promise on the degrees, using only $O(\epsilon^{-1})$ adaptive queries or $O(\epsilon^{-2+1/(\Delta+2)}+\epsilon^{-2+2/W})$ non-adaptive queries, where $\Delta$ and $W$ are parameters depending only on $H$. When $H$ is a simple $t$-cycle, $\Delta=2$ and $W=t$, and the non-adaptive upper bound reduces to $O(\epsilon^{-2+2/t})$, matching the lower bound in Lemma \ref{lemma:nonadaptive-lower-bound}.

Indeed, it suffices for $G$ to be $O(\epsilon)$-close to satisfying the promise. Since \cite{benefits-of-adaptivity} gives an efficient non-adaptive tester for the property of having maximum degree $O(\epsilon N)$ (\cite{benefits-of-adaptivity}), we obtain the following two theorems.

\begin{theorem}[Adaptive Tester]\label{theorem:adaptive-testing-lowdegree-blowup-collections}
For all graphs $H$ and constants $c>1$, there exists an adaptive property tester (with two-sided error) for the (proximity-dependent) combined property of having maximum degree $c\epsilon N$ and being a blow-up collection of $H$. The tester has query complexity $O(\epsilon^{-1}\lg^3{\epsilon^{-1}})$.
\end{theorem}
Since any tester must make $\Omega(\epsilon^{-1})$ queries, Theorem \ref{theorem:adaptive-testing-lowdegree-blowup-collections} is optimal up to a polylogarithmic factor. 

Theorem \ref{theorem:adaptive-testing-lowdegree-blowup-collections}, combined with Lemma \ref{lemma:nonadaptive-lower-bound}, suffices to show a gap of size $Q=\Omega(q^{-2+\delta})$ for all $\delta>0$. In the following theorem, we strengthen this result by proving a tight upper-bound on the non-adaptive query complexity. This shows that there is an infinite hierarchy of achievable relationships between the adaptive and non-adaptive query complexities of proximity-dependent properties.

\begin{theorem}[Non-Adaptive Tester]\label{theorem:nonadaptive-testing-lowdegree-blowup-collections}
For all graphs $H$ and constants $c>1$, there exists a non-adaptive property tester (with two-sided error) for the (proximity-dependent) combined property of having maximum degree $c\epsilon N$ and being a blow-up collection of $H$. The tester has query complexity $O(\epsilon^{-2+1/(\Delta+2)}+\epsilon^{-2+2/W})$, where
$\Delta=\deg(H)$ is the maximum degree of $H$ and $W<\abs{H}^2$ is a bound on the size of a witness against $H$ (see Definition \ref{definition:minimal-witness}).
\end{theorem}

As mentioned previously, when $H$ is a simple $t$-cycle, $\Delta=2$ and $W=t$. Therefore, combining Theorems \ref{theorem:adaptive-testing-lowdegree-blowup-collections} and \ref{theorem:nonadaptive-testing-lowdegree-blowup-collections} with Lemma \ref{lemma:nonadaptive-lower-bound} yields Theorem \ref{theorem:hierarchy-theorem}.


We note that the algorithms in both theorems have running time polynomial in the query complexity. In the adaptive case, the query complexity can also be made polynomial in $c$ and $\abs{H}$. However, we do not do so here, opting instead to simplify the proof.

\subsection{Organization of The Paper}
Section \ref{section:notation} contains basic definitions and notation. In section \ref{section:adaptive-tester}, we show how to adaptively test whether a graph is a blow-up collection, given the promise that the graph is close to having maximum degree $c\epsilon N$. In section 4, we present the non-adaptive algorithm, under the same promise. Finally, section 5 removes the promise by explicitly testing that the input is close to having low degree.

\vspace{-0.5em}\section{Notation and Basics}\label{section:notation}\vspace{-0.5em}

All graphs are assumed to be undirected. Following standard graph-theoretic notation, we let $\Gamma(v)=\{u\st(u,v)\in E\}$ denote the neighbors of $v$. Given $S\subset V$, we let $G|_S$ denote the subgraph induced by $S$ and $\Gamma_S(v)=\Gamma(v)\cap S$ denote the neighbors of $v$ in $S$. Given $S,T\subset V$, we let $E(S,T)=\{(u,v)\in E\st u\in S, v\in T\}$ denote the set of edges between $S$ and $T$ and $S\symdiff T$ denote the symmetric difference of $S$ and $T$.

\begin{definition}[Graph Blow-Up]
A graph $G=([N],E)$ is a \emph{blow-up} of the graph $H=([h],F)$ if there exists a partition of $[N]$ into $V_1\cup\ldots\cup V_h$ such that for every $i,j\in[h]$ and $(u,v)\in V_i\times V_j$, $(u,v)\in E$ if and only if $(i,j)\in F$. We denote the set of blow-ups of $H$ by $\mathcal{BU}(H)$.
\end{definition}
Note that no requirement is made as to the relative sizes of the $V_j$. In particular, we allow the case where $\abs{V_j}=0$. 

\begin{definition}[Blow-Up Collection]
A graph $G=([N],E)$ is a \emph{blow-up collection} of the graph $H$ if there exists a partition of $[N]$ into $V^1\cup\ldots\cup V^k$, for some $k$, such that $G|_{V^i}\in\mathcal{BU}(H)$ for all $i$ and $E(V^i,V^j)=\emptyset$ for all $i\not=j$. We denote the set of blow-up collections of $H$ by $\mathcal{BUC}(H)$.
\end{definition}

Throughout this paper, $H$ will be an arbitrary fixed graph and $c>1$ will be an arbitrary fixed constant. We let $h=\abs{H}$, $\Delta=\deg(H)$, and $\mathcal{LD}_{c\epsilon}=\{G\st\deg(G)\leq c\epsilon N\}$.

To prove Theorems \ref{theorem:adaptive-testing-lowdegree-blowup-collections} and \ref{theorem:nonadaptive-testing-lowdegree-blowup-collections}, we will repeatedly use the concept of a set being \emph{$(k,\alpha)$-partitionable}. Informally, a vertex is $(k,\alpha)$-partitionable if almost all of its neighbors can be partitioned into $k$ groups, such that all of the vertices within a part have essentially the same neighbors in $G$. The formal definition is given in Definition \ref{k-alpha-partitionable}, after we introduce notation for such partitions.

\begin{definition}
Given $v\in V$ and $\alpha>0$, let $C_{v,\alpha}(u)=\{w\in \Gamma(v)\st\abs{\Gamma(w)\symdiff\Gamma(u)}$ $< \alpha\epsilon N\}$.
\end{definition}
In other words, $C_{v,\alpha}(u)$ consists off all vertices in $\Gamma(v)$ which have essentially the same neighbors as $u$. When $v$ and $\alpha$ are clear from context, we will omit the subscripts and write $C(u)$ for $C_{v,\alpha}(u)$.

\begin{definition}[$(k,\alpha)$-Partitionable]\label{k-alpha-partitionable}
A vertex $v$ is \emph{$(k,\alpha)$-partitionable} if there exist representatives $u_1,\ldots,u_k\in \Gamma(v)$ such that $\abs{\bigcup_{i=1}^k{C_{v,\alpha}(u_i)}}\geq\abs{\Gamma(v)}-\alpha\epsilon N$.
\end{definition}
Note that if $G\in\mathcal{BUC}(H)$, then every vertex is $(\deg(H),0)$-partitionable.

Finally, we need the idea of a minimal witness against $\mathcal{BUC}(H)$. Informally, a minimal witness is a set of vertices such that the induced subgraph proves that $G$ is not a valid blow-up collection, while any subset does not suffice.
\begin{definition}[Minimal Witness]\label{definition:minimal-witness}
A set $S\subseteq V$ is a \emph{minimal witness} against $\mathcal{BUC}(H)$ if $G|_S\not\in\mathcal{BUC}(H)$, but for any $S'\subsetneq S$, $G|_{S'}\in\mathcal{BUC}(H)$.
\end{definition}

Note that $G|_S\not\in\mathcal{BUC}(H)$ implies that $G\not\in\mathcal{BUC}(H)$. Furthermore, any minimal witness must be connected, since we could otherwise replace the witness with one of its connected components. 

Given $H$, we let $W=W(H)$ denote the maximum size of a minimal witness. It is easy to see that $W<\frac{1}{2}h^2$. When $H$ is a simple $t$-cycle, $t\geq 4$, $W=\abs{H}$.

\section{Adaptively Testing $\mathcal{BUC}(H)$ Given a Promise}\label{section:adaptive-tester}

We begin by showing how to adaptively test whether a graph $G$ is in $\mathcal{BUC}(H)$, given the promise that $G$ is $O(\epsilon)$-close to $\mathcal{LD}_{c\epsilon}$. 

\begin{figure}[p]
 \centering
  \subfloat[
A witness against $v$ being $(\Delta,0)$-partitionable consists of two sets of vertices. The first set consists of $\Delta+1$ vertices $u_1,\ldots,u_{\Delta+1}\in\Gamma(v)$. The second set consists of $\binom{\Delta+1}{2}$ vertices $w_{ij}$ such that $w_{ij}\in\Gamma(u_i)\symdiff\Gamma(u_j)$ for each $i\not=j$. When $v$ is not $(\Delta,\alpha)$-partitionable, there exist $\alpha\epsilon N$ choices for each $u_i$ such that $u_i\not\in C_{v,\alpha}(u_j)$ for any $j\not=i$, so Algorithm \ref{algorithm:adaptive-buc} can find such vertices efficiently. Having chosen such a set, there exist $\alpha\epsilon N$ choices for $w_{ij}$ for each $i\not=j$, allowing algorithm \ref{algorithm:adaptive-buc} to complete the witness efficiently. Steps \ref{adaptive-buc:begin-degree-test}-\ref{adaptive-buc:end-degree-test} search for witnesses of this type.
	]{\label{fig:degree-witness}\includegraphics[width=0.9\textwidth]{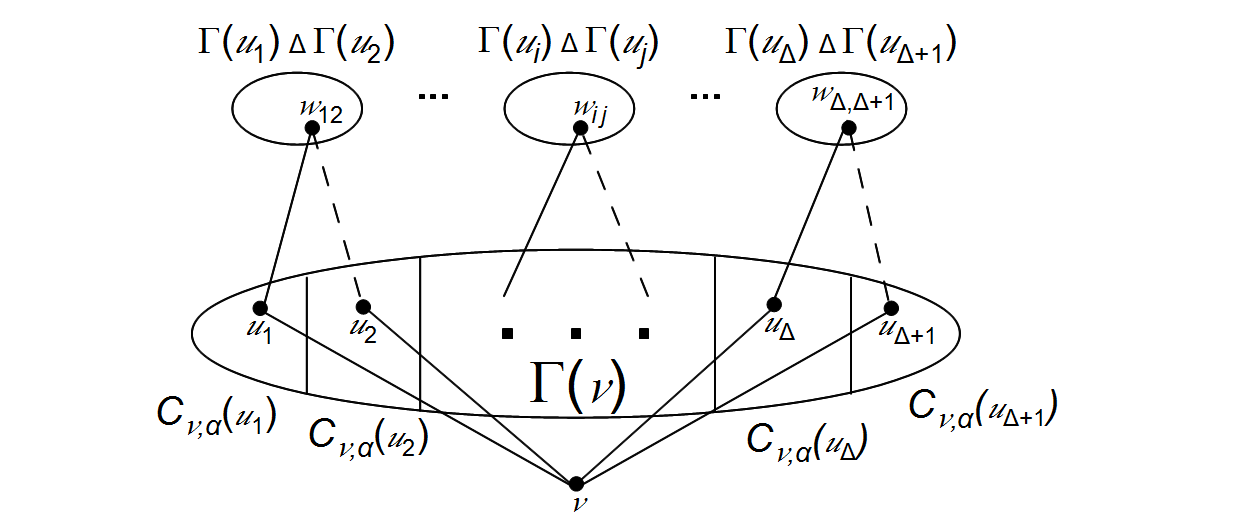}}
	\vspace{0.25in}
  \subfloat[
If $G$ has not been rejected by the first half of Algorithm \ref{algorithm:adaptive-buc}, then $V$ can be partitioned into clusters $V^i_j$, such that each pair of clusters either nearly forms a complete bipartite graph or is nearly disjoint. A minimal witness against having the correct high-level structure consists of a representative vertex from each set such that the induced subgraph is inconsistent with $\mathcal{BUC}(H)$.  For example, when $H=K_4$ is the complete graph on $4$ vertices, the four vertices shown here form a minimal witness. Steps \ref{adaptive-buc:begin-structure-test}-\ref{adaptive-buc:end-structure-test} search for witnesses of this type.
	]{\label{fig:structure-witness}\includegraphics[width=0.9\textwidth]{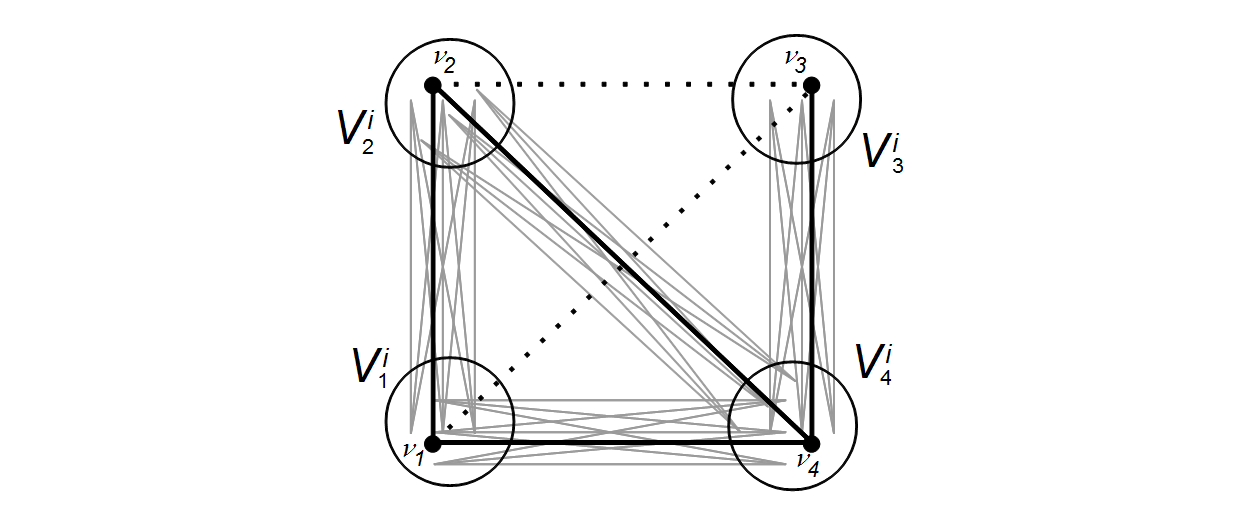}}
  
  \caption{The two types of witnesses used by Algorithm \ref{algorithm:adaptive-buc} to determine whether $G$ is a member of $\mathcal{BUC}(H)$. In each diagram, a line indicates that the algorithm queried that edge, with a solid line indicating the presence of an edge and a dashed line indicating the absence of an edge.}
  \label{fig:witnesses}
\end{figure}

The algorithm consists of two stages. The first stage (steps \ref{adaptive-buc:begin-degree-test}-\ref{adaptive-buc:end-degree-test}) tests whether most vertices are $(\Delta,0)$-partitionable, where $\Delta=\deg(H)$. To do so, it repeatedly selects a vertex and then attempts to find $\Delta+1$ neighbors of that vertex which have mutually distinct neighborhoods (see Figure \ref{fig:witnesses}(a)). If it finds such a set of vertices, they act as a witness against being in $\mathcal{BUC}(H)$ and the algorithm can reject with certainty in step \ref{adaptive-buc:check-degree}. 

If the algorithm fails to find such a witness, then most vertices must be $(\Delta,\alpha)$-partitionable, where $\alpha$ is a small constant. This implies that the graph must be close to being a blow-up of some base graph. In other words, almost all of the vertices can be clustered such that each pair of clusters is either almost disjoint or almost forms a complete bipartite graph. The second stage of the algorithm (steps \ref{adaptive-buc:begin-structure-test}-\ref{adaptive-buc:end-structure-test}) checks if this high-level structure is consistent with $\mathcal{BUC}(H)$ (see Figure \ref{fig:witnesses}(b)). It does so by performing a random search in $G$ for $W=W(H)$ steps, where each step selects a random neighbor of the previously selected vertices. If the resulting $W$ vertices form a witness against $\mathcal{BUC}(H)$, the algorithm rejects in step \ref{adaptive-buc:check-witness}.

\begin{algorithm}\label{algorithm:adaptive-buc}
\begin{turing}{AdaptiveBlowUpCollectionTest$_{H,c}$}{$G;\epsilon$}
\STATE Let $\Delta=\deg(H)$, $W=W(H)$, and $\alpha=(16\Delta\abs{H}^2)^{-1}$. \\~\\
\hspace{-1em}$\backslash\backslash$ \textit{Test whether most vertices are $(\Delta,0)$-partitionable.}
\FOR{$O(c)$ iterations}\label{adaptive-buc:begin-degree-test}
	\STATE Select a random vertex $v$. \label{adaptive-buc:pick-test-vertex}
	\STATE Select a random set $S$ of $O(\Delta\alpha^{-1}\epsilon^{-1})$ vertices, and query $v\times S$. 
	\STATE Let $\overline{S}$ be a random subset of $\Gamma_S(v)$ of size (at most) $c\Delta\alpha^{-1}$.
  \STATE Select a random set $T$ of $O(\Delta^2\alpha^{-1}\epsilon^{-1})$ vertices, and query $\overline{S}\times T$.
  \STATE If the current view of $G$ is inconsistent with $\mathcal{BUC}(H)$, \textsc{Reject}. Specifically, reject if there exist $u_1,\ldots,u_{\Delta+1}\in\overline{S}$ such that $\Gamma_T(u_i)\not=\Gamma_T(u_j)$ for all $i,j$. \label{adaptive-buc:check-degree}
\ENDFOR \label{adaptive-buc:end-degree-test} \\~\\
\hspace{-1em}$\backslash\backslash$ \textit{If $G$ has not yet been rejected, most vertices must be $(\Delta,\alpha)$-partitionable.} \\
\hspace{-1em}$\backslash\backslash$ \textit{Test that $G$ has a high-level structure consistent with $H$.}
\FOR{$O(Wc)^{O(W)}$ iterations}\label{adaptive-buc:begin-structure-test}
  \STATE Select $v_1$ at random, and let $U=\{v_1\}$. \label{adaptive-buc:begin-find-witness}
  \FOR{$j = 2$ \TO $W$}
  	\STATE Select a random set $T_j$ of $O(\Delta\epsilon^{-1})$ vertices, and query $U\times T_j$.
  	\STATE If $\Gamma_{T_j}(U)=\emptyset$, \BREAK. Otherwise, randomly select $v_j\in\Gamma_{T_j}(U)$ and let $U=U\cup\{v_j\}$.\label{adaptive-buc:extend-witness}
  \ENDFOR \label{adaptive-buc:end-find-witness}
  \STATE If $G|_U\not\in\mathcal{BUC}(H)$, \textsc{Reject}.\label{adaptive-buc:check-witness}
\ENDFOR\label{adaptive-buc:end-structure-test}
\STATE If the algorithm hasn't yet rejected, \textsc{Accept}.
\end{turing}
\end{algorithm}

Algorithm \ref{algorithm:adaptive-buc} has query complexity $O(\epsilon^{-1})$, as desired. Furthermore,  it only rejects if it finds a witness against $\mathcal{BUC}(H)$, so the algorithm accepts valid blow-up collections with probability $1$. The following lemma asserts that it rejects graphs which are far from $\mathcal{BUC}(H)$ with high probability.
\begin{lemma}\label{lemma:adaptive-correctness}
Let $G$ be $\frac{\epsilon}{8}$-close to $\mathcal{LD}_{c\epsilon}$ and $\epsilon$-far from $\mathcal{BUC}(H)$. Then Algorithm \ref{algorithm:adaptive-buc} rejects with probability at least $2/3$.
\end{lemma}

The proof is the content of section \ref{section:adaptive-correctness}.

\subsection{Proof of Lemma \ref{lemma:adaptive-correctness}}\label{section:adaptive-correctness}
We first note that if $G$ contains many vertices which are not $(\Delta,\alpha)$-partitionable, then $G$ is rejected with high probability.

\begin{lemma}\label{lemma:adaptive-degree-test}
Let $\Delta=\deg(H)$, $W=W(H)$, and $\alpha=(16\Delta\abs{H}^2)^{-1}$. Suppose that $G$ is $\frac{\epsilon}{8}$-close to $\mathcal{LD}_{c\epsilon}$ and contains at least $\frac{1}{4c}N$ vertices which are not $(\Delta,\alpha)$-partitionable. Then Algorithm \ref{algorithm:adaptive-buc} rejects in step \ref{adaptive-buc:check-degree} with probability at least $2/3$.
\end{lemma}

The proof follows immediately from the definition of $(\Delta,\alpha)$-partitionability, and is given in appendix \ref{proof-of-adaptive-degree-test}.

\vspace{1em}

We now show that if $G$ has at most $\frac{1}{4c}N$ vertices which are not $(\Delta,\alpha)$-partitionable, then $G$ can be partitioned into components such that, for each pair of components, either the edges between them almost form a complete bipartite graph or the components are almost disjoint.

\begin{lemma}\label{lemma:blowup-construction}
Let $\Delta=\deg(H)$ and $\alpha=(16\Delta\abs{H}^2)^{-1}$. Suppose that $G$ is $\frac{\epsilon}{8}$-close to $\mathcal{LD}_{c\epsilon}$ and contains at most $\frac{1}{4c}N$ vertices which are not $(\Delta,\alpha)$-partitionable. Then $G$ is $\frac{5\epsilon}{8}$-close to a graph $\widetilde{G}=(V,\widetilde{E})$ for which the following holds: $V$ can be partitioned into $\bigcup_{i,j}{V^i_j}\cup L$ such that
\begin{enumerate}[parsep=0em,itemsep=0.1em,topsep=0.2em]
\item $\widetilde{\Gamma}(v)=\emptyset$ for all $v\in L$.
\item $E(V^i,V^{i'})=\emptyset$ for all $i\not=i'$.
\item For all $i,j$, $\widetilde{\Gamma}(u)=\widetilde{\Gamma}(v)$ for all $u,v\in V^i_j$.
\item $\abs{V^i_j}>\frac{\epsilon}{16\Delta}N$ for all $i,j$.
\end{enumerate}
where $V^i=\bigcup_{j}{V^i_j}$ and $\widetilde{\Gamma}(u)$ is the neighborhood of $u$ in $\widetilde{G}$. Furthermore, \\ $\abs{\Gamma_{V^i}(u)\symdiff\widetilde{\Gamma}_{V^i}(u)}<\alpha\epsilon N$ for all $u\in V^i$.
\end{lemma}
The proof is given in appendix \ref{proof-of-blowup-construction}.

Note that the bound on $\abs{\Gamma_{V^i}(u)\symdiff\widetilde{\Gamma}_{V^i}(u)}$ in the lemma implies that $\widetilde{G}$ is $\frac{\epsilon}{8}$-close to having maximum degree $2c\epsilon N$, since $G$ is $\frac{\epsilon}{8}$-close to having maximum degree $c\epsilon N$ and the degree of each vertex increases by at most $\alpha\epsilon N$ when going from $G$ to $\widetilde{G}$. Therefore, conditions 3 and 4 imply that most of the components $V^i_j$ are connected to at most $32c\Delta$ other components and have size at most $2c\epsilon N$.

Intuitively, these conditions allow us to view each cluster as a supernode and each bipartite graph as an edge. From this viewpoint, $\widetilde{G}$ becomes a bounded-degree graph, with maximum degree $32c\Delta$. Since each supernode contains $\Omega(\epsilon N)$ vertices (by condition 4 in the lemma), we can simulate neighbor queries by querying $O(\epsilon^{-1})$ random vertices and choosing a random neighbor. Furthermore, if $\widetilde{G}$ is $\Omega(\epsilon)$-far from $\mathcal{BUC}(H)$ when viewed as a dense graph, the corresponding bounded-degree graph is $\Omega(1)$-far from $\mathcal{BUC}(H)$.

To formalize this intuition, we first show that the set of witnesses against $\widetilde{G}$ covers a constant fraction of $V$.

\begin{lemma}\label{lemma:many-distinct-witnesses}
Let $\widetilde{G}$ be as in Lemma \ref{lemma:blowup-construction}, and suppose that $\widetilde{G}$ is $\frac{3\epsilon}{8}$-far from $\mathcal{BUC}(H)$. Then there exist at least $(16Wc^2\epsilon)^{-1}$ distinct sets $\widetilde{W}_k=\bigcup_{\ell}{V^{i_k}_{j_{k,\ell}}}$ such that $G|_{\widetilde{W}_k}$ is a witness against $\mathcal{BUC}(H)$.
\end{lemma}
The proof is given in appendix \ref{proof-of-many-distinct-witnesses}.

We are now ready to show that the second half of Algorithm \ref{algorithm:adaptive-buc} finds a witness against $G$ with high probability. 

By Lemma \ref{lemma:many-distinct-witnesses}, step \ref{adaptive-buc:begin-find-witness} of the algorithm selects a vertex $v_1\in V^i_{j_1}$ corresponding to a witness with high probability. Having chosen $v_1$, we wish to bound the probability that $v_2,\ldots,v_W$ are chosen so as to form a complete witness. Recalling that any minimal witness is connected, there must exist a $v_2\in V^i_{j_2}$ which extends the witness in $\widetilde{G}$. By condition 3 of Lemma \ref{lemma:blowup-construction}, any vertex in $V^i_{j_2}$ can be used in place of $v_2$ to extend the witness. However, since $\abs{V^i_{j_2}}>\frac{\epsilon}{16\Delta}N$ and $\abs{\Gamma_{V^i}(u)\symdiff\widetilde{\Gamma}_{V^i}(u)}<\alpha\epsilon N < \frac{1}{2W}\abs{V^i_{j_2}}$, at least half of $V^i_{j_2}$ is a valid choice to extend the witness in $G$. Furthermore, most of these must have degree $O(c\epsilon N)$.

Iterating this procedure, we see that there are always at least $\Omega(\frac{\epsilon}{\Delta}N)$ choices for $v_j$ which extend the witness in $G$. Furthermore, since each previous $v_{j'}$ was chosen to have degree  $O(c\epsilon N)$, there are at most $O(Wc\epsilon N)$ potential choices for $v_j$. Therefore, with probability $\Omega((W^2c)^{-1})$, Algorithm \ref{algorithm:adaptive-buc} chooses a good $v_j$ at each step and finds a complete witness in $G$.

The formal proof is given in Appendix \ref{proof-of-adaptive-correctness}.

\section{Non-adaptively Testing $\mathcal{BUC}(H)$ Given a Promise}

We now show how to non-adaptively test whether a graph is in $\mathcal{BUC}(H)$, given the promise that $G$ is $O(\epsilon)$-close to $\mathcal{LD}_{c\epsilon}$. 

As in the adaptive case, the first stage of the algorithm verifies that most vertices are $(\Delta,\alpha)$-partitionable. Assuming the graph passes the first stage, the second stage checks that the high-level structure of $G$ is consistent with $\mathcal{BUC}(H)$.

Since we can no longer adaptively restrict our queries to neighbors of a given vertex, we instead rely on the birthday paradox to achieve sub-quadratic query complexity. Recall that the birthday paradox says that given any distribution over a discrete domain $D$, $O(\abs{D}^{1-1/k})$ samples suffice to obtain a $k$-way collision.

\begin{lemma}[Birthday Paradox]\label{lemma:birthday-paradox}
Let $\mathcal{D}$ be a finite domain, and let $\{\mu_i\}$ be a set of probability distribution over $\mathcal{D}$ with $\mu_i(d)=\Omega(1/\abs{\mathcal{D}})$ for all $i$ and $d\in\mathcal{D}$. Suppose that the $i$-th sample is drawn according to $\mu_i$. Then $O(\abs{\mathcal{D}}^{1-1/k})$ samples suffice to obtain a $k$-way collision with high probability.
\end{lemma}

Very informally, a collision will correspond to choosing multiple vertices from a single witness in such a way that a $W$-way collision corresponds to a complete witness. Assuming that the set of witnesses has size $O(\epsilon^{-1})$ and that we can sample from that set with only constant overhead, the birthday paradox implies that $O(\epsilon^{-1+1/W})$ random vertices suffice to find all $W$ vertices corresponding to a complete witness. Querying the entire induced subgraph then yields the desired result. The primary challenge, therefore, is to show that when $G$ is $\epsilon$-far from $\mathcal{BUC}(H)$, the algorithm can efficiently sample from the space of witnesses.

\begin{algorithm}\label{algorithm:nonadaptive-buc}
\begin{turing}{NonAdaptiveBlowUpCollectionTest$_{H,c}$}{$G;\epsilon$}
\STATE Let $\Delta=\deg(H)$, $W=W(H)$, and $\alpha=(16\Delta\abs{H}^{2})^{-1}$. \\
\hspace{-1em}$\backslash\backslash$ \textit{Check that most vertices are $(\Delta,0)$-partitionable.}
\STATE Select a set $S$ of $O((\alpha\epsilon)^{-1+1/(\Delta+2)})$ random vertices, and query $S\times S$. \label{nonadaptive-buc:begin-test-degree}
\STATE Select a set $T$ of $O(\Delta^2\alpha^{-1}\epsilon^{-1})$ random vertices, and query $S\times T$.
\STATE If the current view of $G$ is inconsistent with $\mathcal{BUC}(H)$, \textsc{Reject}. Specifically, reject if there exist $v\in S$ and $u_1,\ldots,u_{\Delta+1}\in\Gamma_S(v)$ such that $\Gamma_T(u_i)\not=\Gamma_T(u_j)$ for all $i,j$. \label{nonadaptive-buc:check-degree}  \\~\\
\hspace{-1em}$\backslash\backslash$ \textit{Check that $G$ has a high-level structure consistent with $H$.}
\STATE Select a set $S$ of $O((\Delta c^2\epsilon)^{-1+1/W})$ random vertices and query $S\times S$. \label{nonadaptive-buc:find-witness}
\STATE If $G|_S\not\in\mathcal{BUC}(H)$, \textsc{Reject}.\label{nonadaptive-buc:check-witness}
\STATE If the algorithm hasn't yet rejected, \textsc{Accept}.
\end{turing}
\end{algorithm}

Algorithm \ref{algorithm:nonadaptive-buc} has query complexity $O(\epsilon^{-2+1/(\Delta+2)}+\epsilon^{-2+2/W})$, as desired. Since it only rejects if it finds a witness against $G$, it accepts all graphs in $\mathcal{BUC}(H)$ with probability $1$. It remains to show that it rejects graphs which are far from $\mathcal{BUC}(H)$ with high probability.

\begin{lemma}\label{lemma:nonadaptive-correctness}
Let $\alpha=(16\Delta\abs{H}^{-2})^{-1}$, as in Algorithm \ref{algorithm:nonadaptive-buc}, and let $G$ be $\frac{\alpha\epsilon}{16c}$-close to $\mathcal{LD}_{c\epsilon}$ and $\epsilon$-far from $\mathcal{BUC}(H)$. Then Algorithm \ref{algorithm:nonadaptive-buc} rejects with probability at least $2/3$.
\end{lemma}

The proof is given in appendix \ref{proof-of-nonadaptive-correctness}.

\section{Removing the Low-Degree Promise}

We now show how to test the combined property of being a valid blow-up collection and having low-degree.

In \cite{benefits-of-adaptivity}, the authors give an $\~O(\epsilon^{-1})$-query algorithm for testing whether the input has maximum degree $O(\epsilon N)$.

\begin{lemma}[\cite{benefits-of-adaptivity}, Theorem 3]\label{lemma:low-degree-tester}
Fix $c>1$ and $\beta>0$. There exists a non-adaptive tester with query complexity $\~O(\epsilon^{-1})$ and two-sided error which accepts graphs with maximum degree $c\epsilon N$ with probability at least $2/3$ and rejects graphs which are $\beta\epsilon$-far from having maximum degree $c\epsilon N$ with probability at least $2/3$.
\end{lemma}

To test whether $G$ is in $\mathcal{BUC}(H)\cap\mathcal{LD}_{c\epsilon}$, we therefore run the tester from Lemma \ref{lemma:low-degree-tester}, and if it accepts, we then run either the adaptive or non-adaptive tester for $\mathcal{BUC}(H)$.

All that remains is to show that if $G$ is $\epsilon$-far from $\mathcal{BUC}(H)\cap\mathcal{LD}_{c\epsilon}$, then it must be $\Omega(\epsilon)$-far from $\mathcal{BUC}(H)$ or $\Omega(\epsilon)$-far from $\mathcal{LD}_{c\epsilon}$.

\begin{lemma}\label{distance-to-low-degree-and-buc}
Suppose that $G$ is $\frac{\epsilon}{18c\Delta^2}$-close to $\mathcal{LD}$ and $\frac{\epsilon}{3}$-close to $\mathcal{BUC}(H)$. Then $G$ is $\epsilon$-close to $\mathcal{LD}\cap\mathcal{BUC}(H)$.
\end{lemma}
The proof of Lemma \ref{distance-to-low-degree-and-buc} is given in appendix \ref{proof:distance-to-low-degree-and-buc}.

We now prove Theorems \ref{theorem:adaptive-testing-lowdegree-blowup-collections} and \ref{theorem:nonadaptive-testing-lowdegree-blowup-collections}.

\begin{proof}[Theorems \ref{theorem:adaptive-testing-lowdegree-blowup-collections} and \ref{theorem:nonadaptive-testing-lowdegree-blowup-collections}]
We first run the tester from Lemma \ref{lemma:low-degree-tester} with $\beta=(18c\Delta(H)^2)^{-1}$. If it accepts, we then run Algorithm \ref{algorithm:adaptive-buc} (in the adaptive case) or Algorithm \ref{algorithm:nonadaptive-buc} (in the non-adaptive case) with tolerance $\frac{\epsilon}{3}$.

If $G\in\mathcal{LD}_{c\epsilon}\cap\mathcal{BUC}(H)$, then the low-degree tester accepts with probability at $2/3$ and blow-up collection tester accepts with probability $1$. So $G$ is accepted with probability at least $2/3$.

Suppose that $G$ is $\epsilon$-far from $\mathcal{LD}\cap\mathcal{BUC}(H)$. By Lemma \ref{distance-to-low-degree-and-buc}, either $G$ is
$\frac{\epsilon}{18c\Delta^2}$-far from $\mathcal{LD}$ or $G$ is $\frac{\epsilon}{3}$-far from $\mathcal{BUC}(H)$. In the first case, the low-degree tester rejects with probability at least $2/3$. In the second case, the blow-up collection tester rejects with probability at least $2/3$, by Lemma \ref{lemma:adaptive-correctness} (in the adaptive case) or Lemma \ref{lemma:nonadaptive-correctness} (in the non-adaptive case).\qed
\end{proof}

\section{Conclusions}

We have shown that there exist proximity-dependent graph properties for which a non-adaptive tester must suffer an almost-quadratic increase in its query complexity over an adaptive tester. This shows that the canonical transformation is essentially optimal, in the worst case.

The primary open question is to remove the proximity-dependence from the graph properties used in Theorem \ref{theorem:hierarchy-theorem}. In particular, for any $\delta>0$, does there exist a \emph{single} graph property $\Pi$ such that testing $\Pi$ requires $Q=\Omega(q^{2-\delta})$ queries for all $\epsilon>0$? It also remains to show whether there exists a nearly-quadratic separation when the adaptive algorithm is only allowed one-sided error. One approach to both of these questions is to prove an $\~O(\epsilon^{-1})$ upper-bound for the adaptive query complexity of $\mathcal{BUC}(H)$ for general graphs.

Finally, we reiterate the intriguing question raised in \cite{algorithmic-aspects} as to what relationships are possible between the adaptive and non-adaptive query complexities. Specifically, do there exists properties such that $Q=\~\Theta(q^{2-\delta})$, with $\delta\not=\frac{2}{t}$? In particular, is it true that $Q$ must either be $\~\Theta(q)$ or $\~\Omega(q^{4/3})$?

\paragraph{Acknowledgments:} Thank you to Oded Goldreich and Chris Umans for very helpful comments and discussions about early versions of this work.
\vspace{-1em}

\bibliographystyle{splncs03}
\bibliography{../bibliography}

\newpage\appendix
\section*{Appendix}

\section{Formal Proofs for Section \ref{section:adaptive-tester}}

\subsection{Proof of Lemma \ref{lemma:adaptive-degree-test}}\label{proof-of-adaptive-degree-test}

Consider the $\frac{1}{4c}N$ vertices which are not $(\Delta,\alpha)$-partitionable. By assumption, at most $\frac{1}{8c}N$ of them have degree greater than $3c\epsilon N$. For otherwise, $G$ would have distance at least $\frac{1}{2}\cdot 2c\epsilon N\cdot\frac{1}{8c}N = \frac{\epsilon}{8} N^2$ from $\mathcal{LD}$.

Therefore, with high probability, some iteration chooses a vertex $v$ in step \ref{adaptive-buc:pick-test-vertex} that is not $(\Delta,\alpha)$-partitionable and has degree at most $3c\epsilon N$. Consider that iteration. Note that since $v$ is not $(\Delta,\alpha)$-partitionable, $\abs{\Gamma(v)}>\alpha\epsilon N$.

We view $S$ as $S_1\cup\ldots\cup S_{\Delta+1}$, with $\abs{S_i}=\alpha^{-1}\epsilon^{-1}$. With high probability, $\Gamma_{S_1}(v)\not=\emptyset$, so let $u_1$ be an arbitrary vertex in $\Gamma_{S_1}(v)$. Since $v$ is not $(\Delta,\alpha)$-partitionable, $\Gamma(v)$ contains at least $\alpha\epsilon N$ vertices $u'$ such that $\abs{\Gamma(u_1)\symdiff\Gamma(u')}>\alpha\epsilon N$, and so with high probability $\Gamma_{S_2}(v)$ contains such a vertex. Let $u_2$ be that vertex. We continue in this manner, selecting vertices $u_3,\ldots,u_{\Delta+1}$ such that $\abs{\Gamma(u_i)\symdiff\Gamma(u_j)}>\alpha\epsilon N$ for all $i\not=j$.

If $\abs{\Gamma_S(v)}<c\Delta\alpha^{-1}$, then $\overline{S}=\Gamma_S(v)$. Otherwise, the probability that each vertex in $\overline{S}$ is a valid choice for $u_i$, given $u_1,\ldots,u_{i-1}$, is at least $\frac{\alpha}{3c}$, so with high probability $\overline{S}$ contains the desired $\Delta+1$ vertices.

Viewing $T$ as $\bigcup_{i,j=1}^{\Delta+1}{T_{ij}}$, we see that, with high probability, $T$ contains a vertex in $\Gamma(u_i)\symdiff\Gamma(u_j)$ for all $i\not=j$. The resulting view of $G$ is therefore inconsistent with $\mathcal{BUC}(H)$ and the algorithm rejects, as desired.\qed

\subsection{Proof of Lemma \ref{lemma:blowup-construction}}\label{proof-of-blowup-construction}

First, delete all edges adjacent to any vertex which is not $(\Delta,\alpha)$-partitionable and add those vertices to $L$. The total cost of doing so is at most $\frac{1}{4c}N\cdot c\epsilon N + \frac{\epsilon}{8}N^2 = \frac{3\epsilon}{8}N^2$, since $G$ is $\frac{\epsilon}{8}$ close to $\mathcal{LD}_{c\epsilon}$.

The construction now proceeds in stages, with each stage constructing a $V^i=\bigcup_{j}{V^i_j}$ such that $\widetilde{E}(V^i,V\backslash V^i)=\emptyset$.

To construct $V^1=\bigcup_j{V^1_j}$, choose an arbitrary vertex $v$ with $\abs{\Gamma(v)}>\frac{\epsilon}{8}N$ and $(\Delta,\alpha)$-partition $\Gamma(v)$ into
$\bigcup_{j=1}^{\Delta}{C(u_j)}$. For any $j$ such that $\abs{C(u_j)}>\frac{\epsilon}{16\Delta}N$, add $C(u_j)$ to $V^1$ as a distinct $V^i_j$. Note that since $\abs{\Gamma(v)}>\frac{\epsilon}{8}N$, there must be at least one partition of that size.

While there exists a $u_j$ in $V^1$ such that $\abs{\Gamma(u_j)\backslash V^1}>\frac{\epsilon}{8}N$, choose such a vertex. For consistency of notation, we relabel $u_j$ to $v$. Let $\bigcup_{j=1}^{\Delta}{C(u_{j})}$ be a $(\Delta,\alpha)$-partitioning of $\Gamma(v)$. Again, there must be a component such that $\abs{C(u_j)\backslash V^1}>\frac{\epsilon}{16\Delta}N$. We add those $C(u_j)\backslash V^1$ to $V^1$ as distinct $V^i_j$s.

Once every $u_j$ in $V^1$ has $\abs{\Gamma(u_j)\backslash V^1}<\frac{\epsilon}{8}N$, we are ready to finalize $V^1$. We begin by separating $V^1$ from $V\backslash V^1$ by deleting $E(V^1,V\backslash V^1)$, at a cost per vertex of at most $\frac{\epsilon}{8}N+\alpha\epsilon N$. We next ensure that $E(V^1_j,V^1_{j'})$ is either $V^1_j\times V^1_{j'}$ or the empty set, for all $j,j'$. This can be easily shown to require at most $\alpha\epsilon N$ edits per vertex.

Note that $V^1$ now satisfies conditions 2-5, by construction.

We now repeat this entire process on the remaining vertices, creating $V^2$, $V^3$, $\ldots$, until every vertex in $V$ has either been covered or has degree at most $\frac{\epsilon}{8}N$. Finally, we delete all edges adjacent to the leftover vertices, at a cost of $\frac{\epsilon}{8}N$ per vertex, and add those vertices to $L$.

The total cost of this procedure is bounded by $(\frac{\epsilon}{8}N+2\alpha\epsilon N)N<\frac{\epsilon}{4}N^2$, for a final cost of $\frac{5\epsilon}{8}N^2$, as desired.\qed

\subsection{Proof of Lemma \ref{lemma:many-distinct-witnesses}}\label{proof-of-many-distinct-witnesses}

First, delete $\frac{\epsilon}{8}N^2$ edges so that $\widetilde{G}$ is in $\mathcal{LD}_{2c\epsilon}$. Note that the resulting graph is still $\frac{\epsilon}{4}$-far from $\mathcal{BUC}(H)$.

Consider some $\widetilde{W}_k$. Completely disconnecting all vertices contained in $\widetilde{W}_k$ requires at most $W(2c\epsilon N)^2$ deletions, since there are at most $W$ components and, by Lemma \ref{lemma:blowup-construction}, each component contains at most $2c\epsilon N$ vertices and each vertex has degree at most $2c\epsilon N$.

Therefore, if there are less than $(16Wc^2\epsilon)^{-1}$ disjoint $\widetilde{W}_k$, the total cost of deleting all witnesses in $\widetilde{G}$ is at most $(16Wc^2\epsilon)^{-1}\cdot W(2c\epsilon N)^2=\frac{\epsilon}{4}N^2$. This contradicts the assumed distance to $\mathcal{BUC}(H)$, and the lemma follows.\qed

\subsection{Proof of Lemma \ref{lemma:adaptive-correctness}}\label{proof-of-adaptive-correctness}

If $G$ is rejected with probability at least $2/3$ by the first part of Algorithm \ref{algorithm:adaptive-buc}, then we are done. So suppose
otherwise. Then by Lemma \ref{lemma:adaptive-degree-test}, there exists $\widetilde{G}$ satisfying the conclusion of Lemma \ref{lemma:blowup-construction}. Since $G$ is $\epsilon$-far from $\mathcal{BUC}(H)$, $\widetilde{G}$ is at least $\frac{3\epsilon}{8}$-far from $\mathcal{BUC}(H)$.

By Lemma \ref{lemma:many-distinct-witnesses}, there exist $(16Wc^2\epsilon)^{-1}$ sets $\widetilde{W}_k=\bigcup_{\ell}{V^{i_k}_{j_{k,\ell}}}$ corresponding to witnesses. Call a $\widetilde{W}_k$ \emph{high-degree} if, for some $V^{i_k}_{j_{k,\ell}}\subset\widetilde{W}_k$ contains $\frac{\epsilon}{64\Delta}N$ vertices of degree greater than $65\Delta Wc^2\epsilon N$. We note that at most half of the $\widetilde{W}_k$ are high-degree. Otherwise, $G$ would have distance at least $\frac{1}{2}\cdot(16Wc^2\epsilon)^{-1}\cdot 64\Delta Wc^2\epsilon N \cdot \frac{\epsilon}{64\Delta}N = \frac{\epsilon}{32}N^2$ to $\mathcal{LD}_{c\epsilon}$, contrary to assumption.

We therefore restrict our attention to the $(32Wc^2\epsilon)^{-1}$ witnesses which are not high-degree.

Let $V^i_1,\ldots,V^i_\ell$, $\ell<W$, be components corresponding to a partial witness in $\widetilde{G}$, and let $v_1,\ldots,v_\ell$ be arbitrary vertices in $V^i_1,\ldots,V^i_\ell$, respectively, such that $\abs{\Gamma(v_j)}<65\Delta Wc^2\epsilon N$ for each $j\leq\ell$. Let $V^i_{\ell+1}$ be a component adjacent to $\bigcup_{j=1}^{\ell}{V^i_j}$ which extends the partial witness in $\widetilde{G}$. Such a component must exist, since witnesses are connected.

Since $\abs{V^i_{\ell+1}}>\frac{\epsilon}{16\Delta}N$, there are at least $\frac{\epsilon}{16\Delta}N$ vertices which progress the witness (in $\widetilde{G}$). Recall that if $V^i_j$ is adjacent to $V^i_{j'}$ in $\widetilde{G}$, then every vertex in $V^i_j$ must be adjacent in $G$ to all but at most $\alpha\epsilon N$ vertices in $V^i_{j'}$ (and similarly if $V^i_j$ is not adjacent to $V^i_{j'}$). Therefore, since $W<\frac{1}{2}\abs{H}^2$ and $\alpha=(16\Delta\abs{H}^2)^{-1}$, there must be at least $\frac{\epsilon}{16\Delta}N-W\alpha\epsilon N > \frac{\epsilon}{32\Delta}N$ choices for $v_{\ell+1}\in V^i_{\ell+1}$ such that $E(v_j,v_{\ell+1})=\widetilde{E}(v_j,v_{\ell+1})$ for all $j\leq \ell$, which means that $v_{\ell+1}$ progresses the witness in $G$ as well. Of these, at least $\frac{\epsilon}{64\Delta}N$ must also have degree at most $65\Delta Wc^2\epsilon N$.

Since $\abs{T_i}=O(\Delta^{-1}\epsilon^{-1})$, with high probability $\Gamma_{T_i}(U)$ contains such a vertex in step \ref{adaptive-buc:extend-witness}. Since each of the $\ell$ vertices selected so far is adjacent to at most $65c\epsilon N$ vertices, there are at most $\ell 65c\epsilon N$ candidates for $v_{\ell+1}$, and so the probability that the selected $v_{\ell+1}$ progresses the witness in $G$ is at least $\frac{\epsilon/32\Delta}{\ell 65c\epsilon} =\Omega((W^2c)^{-1})$.

Therefore, with high probability, the algorithm finds a complete witness in some iteration of steps \ref{adaptive-buc:begin-find-witness}-\ref{adaptive-buc:end-find-witness} and rejects $G$.\qed

\section{Proof of Correctness for the Non-Adaptive Tester}\label{proof-of-nonadaptive-correctness}

As in the proof for the adaptive algorithm, we first show that if $G$ contains many vertices which are not $(\Delta,\alpha)$-partitionable, then $G$ is rejected with high probability.

\begin{lemma}\label{lemma:nonadaptive-degree-test}
Let $\Delta=\deg(H)$, $W=W(H)$, and $\alpha=(16\Delta\abs{H}^2)^{-1}$. Suppose that $G$ is $\frac{\alpha\epsilon}{16c}$-close to $\mathcal{LD}_{c\epsilon}$ and contains at least $\frac{1}{4c}N$ vertices which are not $(\Delta,\alpha)$-partitionable. Then Algorithm \ref{algorithm:nonadaptive-buc} rejects in step \ref{nonadaptive-buc:check-degree} with probability at least $2/3$.
\end{lemma}



For ease of exposition, we begin by proving the claim under the stronger assumption that $G\in\mathcal{LD}_{c\epsilon}$. We then show how to modify the argument to only require that $G$ be $O(\epsilon)$-close to $\mathcal{LD}_{c\epsilon}$.

\begin{proof}
We view the algorithm as choosing $S$ one vertex at a time. As the samples are chosen, we maintain an approximate partitioning of $V$ into $\Omega(\epsilon^{-1})$ sets $\bigcup_k{U_k}\cup L$, such that $\abs{U_k}>\alpha\epsilon N$ for all $k$ and no vertex belongs to more than $\Delta$ of the $U_k$. For each $k$, we initially say that $U_k$ is \emph{uninitialized}. The first time that we select $v\in U_k$, we say that $U_k$ is \emph{initialized with seed $v$}, after which we require that $U_k\subseteq\Gamma(v)$.

Initially, let $L$ consist of all $(\Delta,\alpha)$-partitionable vertices. Arbitrarily divide the remaining $\frac{1}{4c}N$ vertices evenly between the $U_k$.

Whenever $u\in U_k$ is chosen, we update the partitioning as follows. If $U_k$ is uninitialized, then $u$ must not be $(\Delta,\alpha)$-partitionable (by construction). We associate $U_k$ with $u$ and set $U_k=\Gamma(u)$. Note that since $u$ is not $(\Delta,\alpha)$-partitionable, $\abs{U_k}>\alpha\epsilon N$ as required. We also update the remaining uninitialized $U_{k'}$ by removing any vertex $v\in U_{k'}$ such that $v\in\Gamma(u)$ or $\abs{\Gamma(u)\symdiff\Gamma(v)}<\frac{1}{2}\alpha\epsilon N$. Finally, we rebalance the sizes of the uninitialized $U_{k'}$. Supposing for the moment that $G\in\mathcal{LD}_{c\epsilon}$, we remove at most $\abs{\Gamma(u)}\cdot c\epsilon N / \frac{1}{2}\alpha\epsilon N < 2c^2\alpha^{-1}\epsilon N$ vertices.

If vertices $v,u_1,\ldots,u_\ell$, $\ell\leq\Delta$, have been chosen from $U_k$, let 
$$U_k=\Gamma(v)\backslash\bigcup_{i=1}^{\ell}{C_{v,\alpha}(u_i)}.$$
Note that this ensures that $u_i\in\Gamma(v)$ and $\abs{\Gamma(u_i)\symdiff\Gamma(u_j)}>\alpha\epsilon N$ for all $i,j$. Furthermore, since $v$ is not $(\Delta,\alpha)$-partitionable, we must have $\abs{U_k}>\alpha\epsilon N$.

We continue this way until either (i) $\Delta+2$ samples have been chosen from a single $U_k$ or (ii) at least half of the vertices in some initialized $U_k$ belong to at least $\Delta$ other initialized $U_{k'}$. We claim that this occurs within $O(\epsilon^{-1+1/(\Delta+2)})$ steps.

To see this, suppose that condition (ii) has not occurred. First note that either $(8c^3\alpha^{-1}\epsilon)^{-1}$ of the $U_k$ have been initialized, or at least $\frac{1}{8c}N$ vertices are contained in the uninitialized $U_k$. Therefore, 
$$\abs{\bigcup_k{U_k}}>(8c^3\alpha^{-1}\epsilon)^{-1}\cdot\frac{\alpha\epsilon}{2\Delta}N = \frac{\alpha^2}{8c^3}N = \Omega(N),$$
so $O(1)$ samples from $V$ suffice to sample from $U_k$. By the birthday paradox, it follows that $O((\frac{\Delta}{\alpha}\epsilon)^{-1+1/(\Delta+2)})=O(\epsilon^{-1+1/(\Delta+2)})$ samples suffice to obtain the desired $(\Delta+2)$-way collision, with high probability.

\vspace{1em}

Suppose that we halted due to having chosen $v,u_1,\ldots,u_{\Delta+1}$ from some $U_k$. Then with high probability, $T$ contains a vertex from $\Gamma(u_i)\symdiff\Gamma(u_j)$, for each $i,j$, and the algorithm rejects in step \ref{nonadaptive-buc:check-degree}.

If, instead, we halted due to condition (ii), then there exist $\frac{\alpha\epsilon}{2}N$ vertices which are adjacent to $\Delta+1$ of the seed vertices. With high probability, $T$ contains such a vertex $u$. Let $v_1,\ldots,v_{\Delta+1}$ be the corresponding seed vertices. Recall that, by construction, $\abs{\Gamma(v_i)\symdiff\Gamma(v_j)}>\frac{1}{2}\alpha\epsilon N$ for all seed vertices $v_i,v_j$. Therefore, with high probability, $T$ contains a vertex in $\Gamma(v_i)\symdiff\Gamma(v_j)$ for each $i\not=j$, and the algorithm rejects in step \ref{nonadaptive-buc:check-degree}.

\vspace{1em}

Note that in the preceding argument, we only assumed that $G\in\mathcal{LD}_{c\epsilon}$ when bounding the number of vertices discarded after each step from the unitialized $U_{k'}$. Note, however, that each additional discarded vertex implies an additional $\frac{1}{2}\alpha\epsilon N$ distance from $\mathcal{LD}_{c\epsilon}$. Therefore, we delete at most an additional $\frac{1}{16c}N$ vertices. It follows that $\abs{\bigcup_k{U_k}}=\Omega(N)$, as required, and the rest of the proof goes through unchanged.\qed

\end{proof}

By Lemma \ref{lemma:nonadaptive-degree-test}, either $G$ is rejected with high probability by step \ref{nonadaptive-buc:check-degree} or we can apply Lemma \ref{lemma:blowup-construction} to obtain $\widetilde{G}$, as in the adaptive case. By Lemma \ref{lemma:many-distinct-witnesses}, it follows that there are many distinct witnesses against $\widetilde{G}$, which allows us to again apply the birthday paradox.

Specifically, let $\mathcal{D}=\{\widetilde{W}_k\}$ be the set of witnesses guaranteed by Lemma \ref{lemma:many-distinct-witnesses}. As was shown in the adaptive case, for any choice of $v_1,\ldots,v_\ell$ corresponding to some $\widetilde{W}_k$, there exist at least $\frac{\epsilon}{32\Delta}N$ choices for $v_{\ell+1}$ which correctly extend $\widetilde{W}_k$ in $G$. Therefore, $O(Wc^2\Delta)$ samples suffice to obtain a vertex which extends some witness. Applying the birthday paradox, we see that step \ref{nonadaptive-buc:find-witness} finds a complete witness with high probability, in which case the algorithm rejects in step \ref{nonadaptive-buc:check-witness}.

We now formalize this argument.

\begin{proof}[Lemma \ref{lemma:nonadaptive-correctness}]

If $G$ is rejected with probability at least $2/3$ by the first part of Algorithm \ref{algorithm:nonadaptive-buc}, then we are done. So suppose otherwise. Then by Lemma \ref{lemma:nonadaptive-degree-test}, there exists $\widetilde{G}$ satisfying the conclusion of Lemma \ref{lemma:blowup-construction}. So by Lemma \ref{lemma:many-distinct-witnesses}, there exists at least $(8Wc^2\epsilon)^{-1}$ distinct $\widetilde{W}_k$.

As in the proof of Lemma \ref{lemma:adaptive-correctness}, let $V^i_1,\ldots,V^i_\ell$, $\ell<W$, be components corresponding to a partial witness in $\widetilde{G}$, and let $v_1,\ldots,v_\ell$ be arbitrary vertices in $V^i_1,\ldots,V^i_\ell$, respectively. Let $V^i_{\ell+1}$ be a component adjacent to $\bigcup_{j=1}^\ell{V^i_j}$ which extends the partial witness in $\widetilde{G}$. Such a component must exist, since minimal witnesses are connected.

Since $\abs{V^i_{\ell+1}}>\frac{\epsilon}{16\Delta}N$, there are at least $\frac{\epsilon}{16\Delta}N$ vertices which progress the witness (in $\widetilde{G}$). Also recall that if $V^i_j$ is adjacent to $V^i_{j'}$ in $\widetilde{G}$, then every vertex in $V^i_j$ must be adjacent in $G$ to all but at most $\frac{\epsilon}{16\Delta h^2}$ vertices in $V^i_{j'}$ (and similarly if $V^i_j$ is not adjacent to $V^{i'}_{j'}$). Therefore, since $W<\frac{1}{2}h^2$, there must be at least $\frac{\epsilon}{16\Delta}N-W\frac{\epsilon}{16\Delta h^2}N > \frac{\epsilon}{32\Delta}N$ choices for $v_{\ell+1}$ such that $E(v_j,v_{\ell+1})=\widetilde{E}(v_j,v_{\ell+1})$ for all $j\leq \ell$, which means that $v_{\ell+1}$ progresses the witness in $G$ as well.

We now map step \ref{nonadaptive-buc:find-witness} onto the birthday paradox. Given vertices $v^k_1,\ldots,v^k_{\ell}$ in $V^{i_k}_{j_{k,1}},\ldots,V^{i_k}_{j_{k,\ell}}$, respectively, let $U_k\subseteq V^{i_k}_{j_{k,\ell+1}}$ be the set of vertices which extend $\widetilde{W}_k$ in $G$.

We let $\mathcal{D}=\{U_k\}$. Since $\abs{U_k}>\frac{\epsilon}{32\Delta}N$ for all $k$ and the $U_k$ are disjoint, $O(Wc^2\Delta)$ samples from $V$ suffice to obtain a sample from $\mathcal{D}$. Therefore, by Lemma \ref{lemma:birthday-paradox}, $O(\epsilon^{-1+1/W})$ samples suffice to obtain a complete witness, which means that $S$ contains a complete witness with high probability and the algorithm rejects in step \ref{nonadaptive-buc:check-witness}.\qed
\end{proof}

\section{Proof of Lemma \ref{distance-to-low-degree-and-buc}}\label{proof:distance-to-low-degree-and-buc}

Let $\epsilon_1=\frac{\epsilon}{18c\Delta^2}$ be the distance to $\mathcal{LD}_{c\epsilon}$ and $\epsilon_2=\frac{\epsilon}{3}$ be the distance to $\mathcal{BUC}(H)$. Let $V=\bigcup{V^i_j}$ be the optimal decomposition of $G$ with respect to $\mathcal{BUC}(H)$. 

First, we completely disconnect all $V^i_j$ such that $\abs{V^i_j}<\frac{\epsilon}{3\Delta}N$. Next, we delete all superfluous edges and add all missing edges so that $G\in\mathcal{BUC}(H)$, at total cost at most $\epsilon_2N^2$. Note that since $\bigcup{V^i_j}$ is the optimal decomposition, every vertex must have been connected to at least half of its neighbors. Therefore, at worst this doubled the degree of every vertex, thereby doubling the distance to $\mathcal{LD}$. Note that $G$ is now a valid blow-up collection.

The total cost of the edits so far is bounded by $\epsilon_2N^2 + \sum{\abs{V^i_j}\Delta\frac{\epsilon}{3\Delta}N} = \epsilon_2N^2 +
\frac{\epsilon}{3}N^2$, since each component is adjacent to at most $\Delta$ other components.

For each $i,j$, if $\abs{V^i_j}>c\epsilon N$, delete $\abs{V^i_j}-c\epsilon N$ vertices from $V^i_j$. This clearly preserves membership in $\mathcal{BUC}(H)$. Furthermore, the cost of doing the deletion is exactly equal to the decrease in distance to $\mathcal{LD}_{c\epsilon}$. To see this, note that every neighbor of $V^i_j$ needed to delete at least that many edges into $V^i_j$, and by symmetry we can assume that they all delete corresponding edges.

Finally, while any high-degree vertices remain, we do the following. First, choose $u\in V^i_j$ such that $\abs{\Gamma(u)}>c\epsilon N$ and $v\in\Gamma(u)$, and delete $v$. Note that the cost of doing so is, at most, $c\epsilon N\cdot\Delta$ and the decrease in distance is at least $\abs{V^i_j}>\frac{\epsilon}{3\Delta}N$, for a net multiplicative overhead of $3c\Delta^2$. 

The total cost of removing the high-degree vertices is therefore bounded by $3c\Delta^2\cdot2\epsilon_1N^2$, which means that the final cost is bounded by
$(\epsilon_2N^2 + \frac{\epsilon}{3}N^2) + (3c\Delta^2\cdot 2\epsilon_1N^2) =
\epsilon N^2$ as desired. \hfill$\qed$

\end{document}